\documentclass[prd,aps,amsmath,twocolumn,superscriptaddress,amssymb,reprint,nofootinbib]{revtex4-1}

\usepackage{graphicx}
\usepackage{color}
\usepackage{amsmath, slashed}
\usepackage{braket}
\usepackage{placeins}

\begin{document}
\title{B-physics anomalies: The bridge between R-parity violating Supersymmetry and flavoured Dark Matter} 

\author{Sokratis Trifinopoulos}
\affiliation{Physik-Institut, Universit\"at Z\"urich, CH-8057 Z\"urich, Switzerland}

\begin{abstract}
In recent years, significant experimental indications that point towards Lepton Flavour Universality violating effects in B-decays, involving $b \to c \tau \nu$ and $b \to s \ell^+ \ell^-$ have been accumulated. A possible New Physics explanation can be sought within the framework of R-parity violating Supersymmetry, which contains the necessary ingredients to explain the anomalies via both leptoquark, tree-level exchange and one-loop diagrams involving purely leptonic interactions. In addition, an approximate $U(2)^2$ flavour symmetry, that respects gauge coupling unification, successfully controls the strength of these interactions. Nevertheless strong constraints from leptonic processes and $Z$ boson decays exclude most of the relevant parameter space at the $2 \sigma$ level. Moreover, R-parity violation deprives Supersymmetry of its Dark Matter candidates. Motivated by these deficiencies, we introduce a new gauge singlet superfield, charged under the flavour symmetry and show that its third-generation, scalar component may participate in loop diagrams that alleviate the above-mentioned tensions, while at the same time reproduce the observed relic abundance. We obtain a solution to both anomalies that is also fully consistent with the rich Flavour and Dark Matter phenomenology. Finally, we assess the prospect to probe the model at future experiments.
\end{abstract}

\maketitle

\section{Introduction}

Testing the limits of the Standard Model (SM) by examining processes that might not respect Lepton Flavor Universality (LFU) is one of the most prominent endeavors to discover New Physics (NP) pursued at the LHC and several other experiments. Intriguingly, recent data exhibit anomalies in rare B-meson LFU violating decays, as encoded by the ratios:
	\begin{align}
	 &R_{K^{(*)}} = \frac{ \mathcal B(B \to K^{(*)} \mu\overline{\mu}) }{ \mathcal B(B \to K^{(*)} e \bar{e} )}, \ \ \
	R_{D^{(*)}} = \frac{ \mathcal B(B \to D^{(*)} \tau \overline{\nu})}{ \mathcal B(B \to D^{(*)} \ell \overline{\nu} ) } \label{eq:Ratios}
	\end{align} 
which are almost free from theoretical uncertainties in hadronic matrix elements. $R_{D^{(*)}}$ concerns an enhancement of the charged-current interaction $b \to c \tau \nu$~\cite{Lees:2013uzd, Hirose:2016wfn, Aaij:2015yra, Aaij:2017deq, Belle:Moriond} with respect to the tree-level induced SM amplitude~\cite{Bernlochner:2017jka}~\cite{Bigi:2017jbd}, whereas $R_{K^{(*)}}$ a deficit in neutral-current transition involving $b \to s \ell^+ \ell^-$~\cite{Aaij:2017vbb}~\cite{Aaij:2019wad} at one-loop level~\cite{Bordone:2016gaq}. They independently differ by approximately $3-4 \sigma$ from their respective SM predictions and along with reported indications towards LFU violation in less theoretically-clean observables, e.g. the angular observable $P_5^\prime$ in the $B \to K^* \mu\overline{\mu}$ decay~\cite{Aaij:2015oid, Wehle:2016yoi, Ciuchini:2015qxb} and the $R_{J / \Psi}$ ratio~\cite{Aaij:2017tyk, Watanabe:2017mip}, constitute a consistent pattern of deviations that has motivated several attempts for a simultaneous explanation~\cite{Bhattacharya:2014wla, Alonso:2015sja, Greljo:2015mma, Calibbi:2015kma, Bauer:2015knc, Fajfer:2015ycq, Barbieri:2015yvd, Das:2016vkr, Hiller:2016kry, Bhattacharya:2016mcc, Boucenna:2016wpr, Buttazzo:2016kid, Boucenna:2016qad, Barbieri:2016las, Becirevic:2016oho, Buttazzo:2017ixm, Bordone:2017anc, Crivellin:2017zlb, Cai:2017wry, Megias:2017ove, DiLuzio:2017vat, Assad:2017iib, Calibbi:2017qbu, Bordone:2017bld, Barbieri:2017tuq, Greljo:2018tuh, Blanke:2018sro, DiLuzio:2018zxy, Marzocca:2018wcf, Kumar:2018kmr, Guadagnoli:2018ojc, Becirevic:2018afm, Bordone:2018nbg, Fornal:2018dqn, Bhattacharya:2019eji, Cornella:2019hct, Biswas:2014gga, Zhu:2016xdg, Deshpand:2016cpw, Das:2017kfo, Altmannshofer:2017poe, Earl:2018snx, Trifinopoulos:2018rna, Hu:2018lmk, Barbieri:2019zdz}. Nevertheless, the theoretical challenge to devise an ultraviolet (UV) complete model that accommodates all other low-energy observables, has proven to be notoriously difficult and to the best of our knowledge, there have only been a handful of proposals in the bibliography so far~\cite{DiLuzio:2017vat, Bordone:2017bld, Barbieri:2017tuq, Marzocca:2018wcf, Becirevic:2018afm, Blanke:2018sro, Trifinopoulos:2018rna}. \par
Considering that the charged-current anomaly involves the tau and both of them the bottom, NP scenarios in which the third-generation SM fermions plays a special role are favoured. Furthermore, model-independent analyses suggest that $R_{K^{(*)}}$ can be resolved by singly modifying the Wilson Coefficient (WC) of the semi-leptonic di-muon vector and axial operators, i.e. $C_9^{\mu}$ and $C_{10}^{\mu}$~\cite{Descotes-Genon:2015uva, Aebischer:2019mlg, Alguero:2019ppt}. In light of these observations, the potential of R-parity violating (RPV) Supersymmetry (SUSY) to provide a comprehensive solution has been studied~\cite{Biswas:2014gga, Zhu:2016xdg, Deshpand:2016cpw, Das:2017kfo, Altmannshofer:2017poe, Earl:2018snx, Trifinopoulos:2018rna, Hu:2018lmk}. The Minimal Supersymmetric Standard Model (MSSM) particle content contains two third-generation, scalar particles, namely the right-handed sbottom $\tilde b_R$ and stau $\tilde \tau_R$, which in most  models of spontaneous SUSY breaking are predicted to be significantly lighter than the first and second generations of superpartners at the electroweak scale~\cite{Martin:1997ns}. If RPV interactions are turned on, $\tilde b_R$ has the right quantum numbers to mediate the required, large NP effects on $R_{D^{(*)}}$ at tree-level, while according to the novel result of~\cite{Trifinopoulos:2018rna}, $\tilde \tau_R$ and $\tilde b_R$ can enter a box diagram, which generates a negative contribution to $C_9^{\mu}$. Additionally, inspired by the corresponding Effective Field Theory (EFT) studies~\cite{Greljo:2015mma, Barbieri:2015yvd, Bordone:2017anc, Buttazzo:2017ixm}, it was shown that an approximate $U(2)^2$ flavour symmetry\footnote{Historically, the $U(2)$ flavour symmetries have strong ties with SUSY as their initial proposal intended to solve the `flavour problem' of the MSSM~\cite{Barbieri:1995uv}~\cite{Barbieri:2011ci}. This is also true for the WIMP paradigm, since the LSP in R-parity conserving SUSY has been cherished as the prototype of a WIMP particle and a point of reference for direct searches.}, acting only on the first two generations, naturally suppresses the RPV couplings within the experimental bounds and still allows for an improvement over the SM fit. However the possibility to completely alleviate the tensions in the ratios \eqref{eq:Ratios} and especially in $R_{K^{(*)}}$ is severely limited by the strict bounds on $Z$ boson decay to leptons and tree-level, Lepton Flavor Violating (LFV) $\tau$ decays. \par
The RPV setup preserves all the attractive features of SUSY, but one, namely the Lightest Supersymmetric Particle (LSP) acting as Dark Matter (DM) candidate, because RPV interactions render it unstable. Therefore, we are compelled to speculate on the existence of new particles and the simplest, most popular example is that of a weakly interacting massive particle (WIMP)\footnote{See footnote 1.}, which is thermally produced in the early universe. Among the numerous WIMP models, those that assume a direct DM coupling to specific SM fermions and hence furnish distinctive collider signatures, are of particular interest. As with the case of RPV interactions, we would like to justify the apparent suppression of DM interactions with the first SM generation, which is most relevant to direct detection experiments. Consequently, we focus on the subclass of models, that explains from a flavour symmetry rationale the preference of DM to couple predominantly to particular flavours of either quark~\cite{Kamenik:2011nb, Kile:2011mn, Batell:2011tc, Bai:2013iqa, Batell:2013zwa, Kumar:2013hfa, DiFranzo:2013vra, Lopez-Honorez:2013wla, Agrawal:2014una, Agrawal:2014aoa, Kilic:2015vka, Blanke:2017tnb, Blanke:2017fum} or leptons~\cite{Agrawal:2011ze, Agrawal:2014ufa, Yu:2014mfa, Bai:2014osa, Chang:2014tea, Kile:2014jea, Hamze:2014wca, Lee:2014rba, Agrawal:2015kje, Chen:2015jkt, Baker:2018uox}, usually in Minimal Flavour Violation (MFV) -type scenarios. Moreover, supersymmetric flavored DM models exhibit an even more restricted structure~\cite{Batell:2013zwa}. \par
As long as the nature of DM remains a mystery, it is important to investigate any possible connection with other anomalous observations. To this end, there have been attempts to link mostly the $R_{K^{(*)}}$ discrepancy with DM~\cite{Sierra:2015fma, Belanger:2015nma, Kawamura:2017ecz, Ko:2017quv, Fuyuto:2017sys, Cline:2017aed, Cline:2017qqu, Azatov:2018kzb, Rocha-Moran:2018jzu, Bernal:2018aon, Darme:2018hqg, Choi:2018stw, Kumar:2018era, Singirala:2018mio, Hati:2018fzc, Falkowski:2018dsl, Okada:2018tgy, Baek:2018aru, Hutauruk:2019crc, Biswas:2019twf, Cerdeno:2019vpd, Baek:2019qte}. In the current work, we address both anomalies in a supersymmetric framework related to a flavoured hidden sector. The new particle content is economical, simply consisting of a gauge singlet, flavour multiplet and a $SU(2)_{\rm L}$ doublet, flavour singlet, chiral superfield, that acts as the mediator of a DM-lepton interaction. In analogy to the assumed MSSM mass spectrum, the third-generation scalar contained in the multiplet is taken to be the lightest component and thus the DM candidate. We revisit the $U(2)^2$ flavour symmetry that controls the strength of both DM and RPV interactions, so that it is compatible with Grand Unification Theories (GUTs). By virtue of the spin and unsuppressed coupling of the DM particle to the $\tau_L$, contributions are generated in one-loop diagrams involving purely leptonic interactions that interfere destructively with the ones generated by the RPV interactions. By exploring the interplay between the rich flavour and DM phenomenology, we determine whether it is possible to invoke an appropriate cancellation mechanism and at the same time reproduce the correct relic abundance for a natural choice of parameters. The performance of the overall fit is subsequently evaluated. Finally, we briefly discuss the implications for future experiments.

\section{Model}
\label{sec:model}

\subsection{R-parity violating interactions}
\label{sec:RPV_interactions}

Let us first review the R-parity odd and gauge-invariant superpotential, composed exclusively of MSSM quark and lepton superfields~\cite{Barbier:2004ez},
\begin{equation}
\label{eq:RPVW}
W_{\text{RPV}} = \frac{1}{2} \lambda_{ijk} L_i L_j E_k^c + \lambda'_{ijk} L_i Q_j D_k^c + \frac{1}{2} \lambda''_{ijk} U_i^c U_j^c D_k^c,
\end{equation}
where there is a summation over the flavour indices $i,j,k = 1,2,3$, and summation over gauge indices is understood. \par
The traditional motivation for R-parity is that it forbids the baryon-number breaking $\lambda''$ couplings and thus ensures proton stability, but this argument is no longer substantial. Notwithstanding, if the MSSM is an effective theory~\cite{Brust:2011tb}, higher-dimensional operators could also induce rapid proton decay and one should rely on different mechanisms, e.g. MFV-type flavour symmetries, to mitigate the effect~\cite{Smith:2008ju}~\cite{Csaki:2011ge}. \par   
As already noted in the introduction, we may treat only the third generation as effectively supersymmetrized, a premise that is also supported by general bottom-up approaches~\cite{Brust:2011tb}~\cite{Papucci:2011wy}. The low-energy spectrum can be further simplified by assuming that the left-handed superpartners are at least an order of magnitude heavier. The trilinear terms associated with the $\lambda$ and $\lambda'$ couplings that are relevant to our discussion and expanded in standard four-component Dirac notation, are then,
\begin{align}
\label{eq:trilinearlambda}
\mathcal{L}_\lambda&= -\frac{1}{2} \lambda_{ij3} \left(\tilde{\tau}_{R}^* \bar{\nu}_{Ri}^c \ell_{Lj} - (i \leftrightarrow j) \right) + \text{h.c.} \\
\label{eq:trilinearlambdaprime}
\mathcal{L}_{\lambda'}&= - \lambda'_{ij3} \left(\tilde{b}_{R}^* \bar{\nu}_{Ri}^c d_{Lj} - \tilde{b}_{R}^* \bar{\ell}_{Ri}^c u_{Lj} \right) + \text{h.c.}
\end{align}
Here, it is also worth mentioning that since direct searches at LHC have been unfruitful so far~\cite{Autermann:2016les}, a `vanilla' MSSM scenario with at least $10\%$ fine-tuning is ruled out~\cite{Strumia:2011dv}. The exploration of non-minimal realisations of SUSY with reduced missing energy signatures, such as RPV SUSY, as viable model building directions is therefore prominent~\cite{Hardy:2013ywa, Arvanitaki:2013yja, Buckley:2016kvr}. As a matter of fact, our scenario is similar to the Effective RPV SUSY model studied in Ref.~\cite{Buckley:2016kvr}, where the first two generations of squarks decouple from the low-energy spectrum. This is the most successful model in evading LHC bounds for a natural parameter space with a messenger scale of SUSY breaking $20~\rm TeV \lesssim \Lambda < 100~\rm TeV$ considered by the authors.

\subsection{Supersymmetric flavoured Dark Matter}
\label{sec:DM_sector}

We extend the matter content of the MSSM introducing a vector-like DM flavour multiplet $X$, $\bar{X}$ and a vector-like mediator $Y$, $\bar{Y}$. The requirement that $X$ is a gauge singlet and couples to the left-handed leptons fixes the quantum numbers of the new superfields. It follows, that the mediator is charged under the MSSM gauge group. The most general superpotential relevant to the new superfields can be expressed as, 
\begin{equation}
\label{eq:DMW}
W_{\text{DM}} = \hat{M}_X X \bar{X} + \hat{M}_Y Y \bar{Y} + \hat{\lambda}_{ij} X_i Y L_j ,
\end{equation}
where $i,j = 1,2,3$ denote again flavour indices. \par
The crucial difference with nonsupersymmetric versions of flavoured DM is that the $\hat{\lambda}$-term is the only interaction that we are permitted to include at renormalizable level. For instance, the so-called Higgs Portal of scalar DM models, i.e. a scalar cubic interaction between $X$ and the Higgs, is absent. This is an important point, given the fact that such an interaction is strongly constrained by direct detection searches~\cite{Escudero:2016gzx} and in non-supersymmetric scalar DM models one must set the coupling arbitrarily to zero. \par
As far as the mass terms are concerned, they are obviously non-holomorphic and indeed their origin lies in terms of the K\"ahler potential that can be associated with the soft SUSY breaking scale in a way analogous to the Giudice-Masiero mechanism~\cite{Giudice:1988yz}. In synergy with non-canonical kinetic terms, which are expanded as functions of flavour spurions, a large mass splitting between $X_3$ and the nearly degenerate $X_1$ and $X_2$ can be obtained~\cite{Batell:2013zwa}. In addition, soft-breaking terms in the hidden sector produce an additional mass splitting between the scalar and fermionic components. As a result, the scalar component $\chi$ of $X_3$ can be the lightest DM state and together with the fermion component $\psi$ of $Y$, they build the term
\begin{equation}
\mathcal{L}_{\hat{\lambda}} = \hat{\lambda}_{3j} \bar{\ell}_{L j} \chi \psi + \text{h.c.},
\label{eeq:trilinearlambdahat}
\end{equation}
which is the interaction term relevant to low-energy phenomenology. \par
Regarding the DM stabilization, we notice that because $X$ is leptophilic and the flavour-breaking sources are not MFV-type, we cannot apply the mechanism of the residual $Z_3$ symmetry that exists for quark-flavoured DM~\cite{Batell:2011tc, Agrawal:2014aoa}. One can envision alternatives in an extended hidden sector with its own gauge symmetries, but this discussion is beyond the scope of the current paper. We shall simply assume a $Z_2$ symmetry, under which $X$ and $Y$ are odd and their direct decay to SM particles is forbidden, as it is customary in simplified models of leptophilic dark matter~\cite{Agrawal:2011ze, Yu:2014mfa, Bai:2014osa, Hamze:2014wca, Lee:2014rba, Agrawal:2015kje, Chen:2015jkt, Baker:2018uox}.

\subsection{Flavour Structure}
\label{sec:U2}

Following EFT approaches~\cite{Greljo:2015mma, Barbieri:2015yvd, Buttazzo:2017ixm, Bordone:2017anc}, a non-abelian flavour symmetry $U(2)_q \times U(2)_\ell$ was employed in Ref.~\cite{Trifinopoulos:2018rna}. Under the assumption that it is spontaneously broken by the vacuum expectation values (VEVs) of two different sets of flavon fields, one quark- and one lepton-flavoured, the observed fermion mass hierarchy and the phenomenologically viable hierarchy of RPV couplings is reproduced~\cite{Bhattacharyya:1998vw}. Nevertheless, such a breaking pattern is against the idea of gauge coupling unification\footnote{Note that the beyond-the-MSSM elements postulated in this work do not spoil gauge coupling unification. In particular, RPV interactions even in the limit where the first two generations are decoupled, do not alter the RG evolution up to a shift of the unified coupling value~\cite{Altmannshofer:2017poe} and if the SM-charged mediator $Y$ is embedded in a complete GUT multiplet, e.g. for $SU(5)$ in an antifundamental $\bar{\textbf{5}}$, then unification is still preserved.}. It is thus suitable to adopt a different version of the $U(2)^2$ flavour symmetry group that commutes with $SU(5)$, which is contained in all GUT groups. In terms of the $\textbf{10}_i (T_i) \oplus \bar{\textbf{5}}_i (\bar{F}_i)$ ($i=1,2,3$) representations of $SU(5)$, the plausible choice is the flavour symmetry $U(2)_{T} \times U(2)_{\bar{F}}$~\cite{Barbieri:2015bda}, under which the matter superfields transform as,
\begin{align}
&(T_1, T_2) \sim (\textbf{2},\textbf{1}), \ \ \  T_3 \sim (\textbf{1},\textbf{1}), \notag \\
&(\bar{F}_1, \bar{F}_2) \sim (\textbf{1},\textbf{2}), \ \ \  \bar{F}_3 \sim (\textbf{1},\textbf{1}).
\end{align}
In a minimal fashion, we assign $U(2)_{\bar{F}}$ charge to the doublet consisting of the first two generations of $X$. \par
The $SU(5)$-invariant Yukawa Lagrangian,
\begin{align}
\label{eq:LYukawa}
\mathcal{L}_Y &= y_t T_3 T_3 H_5 + y_t x_t \textbf{T} \textbf{V}_T T_3 H_5 + \textbf{T} \Delta_T \textbf{T} H_5 + y_b T_3 \bar{F}_3 H_{\bar{5}} \notag \\
&+ y_b x_b \textbf{T} \textbf{V}_T \bar{F}_3 H_{\bar{5}} + \textbf{T} \Delta_{T \times \bar{F}} \bar{\textbf{F}} H_{\bar{5}},
\end{align}
is also kept flavour-invariant by ascribing to the spurions $\textbf{V}_T$, $\Delta_T$ and $\Delta_{T \times \bar{F}}$ the appropriate transformation properties. The masses and the mixings in the low-energy limit are then qualitatively understood\footnote{The correct Yukawa matrices are derived only after further refinement of Eq. \eqref{eq:LYukawa} by including a term of the form $\textbf{T} \Delta_{T \times \bar{F}}' \bar{\textbf{F}} H_{\bar{45}}$ that generates different $\mu - s$ and $e - d$ masses and leads to the Georgi-Jarlskog mass relation~\cite{Georgi:1979df}.} with the following alignment of the spurions in flavour space,
\begin{equation}
\textbf{V}_T = (0 \ \ \epsilon)^T, \ \ \ \ \Delta_T = \left(\begin{matrix}0 & \epsilon' \\ -\epsilon' & \epsilon \rho \end{matrix}\right), \ \ \ \ \Delta_{T \times \bar{F}} = \left(\begin{matrix}0 & \epsilon' \\ -\epsilon' & \epsilon \end{matrix}\right),
\end{equation}
the symmetry breaking parameters $\epsilon \approx 0.025$ and $\epsilon' \approx 0.004$ and the higher order correcting parameter $\rho \approx 0.02$~\cite{Barbieri:1996ww}. \par 
All trilinear terms in the superpotential \eqref{eq:RPVW} and \eqref{eq:DMW} can be converted to holomorphic flavour singlets by appropriately contracting the matter superfields with the above spurions and an additional spurion $\textbf{V}_{\bar{F}}$ transforming as $(\textbf{1},\bar{\textbf{2}})$. In retrospect, we know that the $b \to s \ell^+ \ell^-$ anomalies imply a sizeable coupling $\lambda_{323}$, possibly of order $\mathcal{O}(1)$. Since there is no term of the form $\bar{\textbf{F}} \textbf{V}_{\bar{F}} \bar{F}_3$ in Eq. \eqref{eq:LYukawa}, we have the freedom to write: $\textbf{V}_{\bar{F}} = (0 \ \ \epsilon_{\bar{F}})^T$ with $\epsilon_{\bar{F}} \approx 1$.  As a result of this symmetry and symmetry-breaking ansatz, we achieve a flavour structure similar to the one proposed in~\cite{Trifinopoulos:2018rna}, but this time with a more appealing theoretical justification. Last but not least, a thorough study of the connection between the RPV trilinear and the Yukawa couplings in the context of $SU(5)$ may reveal the conditions under which the relation $\lambda_{323} \approx \lambda'_{233} \approx \lambda'_{333}$ is a prediction of the model~\cite{Bajc:2015zja}.

\begin{table}[t]
\addtolength{\arraycolsep}{3pt}
\renewcommand{\arraystretch}{1.2}
\centering
\begin{tabular}{|c|c|}
\hline
observable & experimental value \\
\hline\hline
$r_{D^{(*)}}$ &  $1.139 \pm 0.053$\footnotemark[5] \\
\hline
$\delta C_9^{\mu} = -\delta C_{10}^{\mu}$ &  $-0.46 \pm 0.10$\cite{Alguero:2019ppt} \\
\hline
$R_{B \to K^{(*)} \nu \bar{\nu}}$ & $<2.7$~\cite{Grygier:2017tzo} \\
\hline
$C_{B_s}$ & $1.070 \pm 0.088$~\cite{Bona:2007vi} \\
\hline
$\phi_{B_s}$ & $(0.054 \pm 0.951)^{\circ}$~\cite{Bona:2007vi} \\
\hline
$\Delta M_{B_d}$ & $(3.327 \pm 0.013) \times 10^{-13} ~\rm GeV$~\cite{Amhis:2014hma} \\
\hline
$\frac{a_{\tau}}{a_{e}}$ & $1.0019 \pm 0.0015$~\cite{Patrignani:2016xqp} \\
\hline
$R_{\tau}^{\tau/\mu}$ & $1.0022 \pm 0.0030$~\cite{Pich:2013lsa} \\
\hline
$\mathcal B(\tau \to \mu \mu \bar{\mu})$ & $<2.1 \times 10^{-8}$~\cite{Patrignani:2016xqp} \\
\hline
$\mathcal B(\tau \to \mu \gamma)$ & $<4.4 \times 10^{-8}$~\cite{Patrignani:2016xqp} \\
\hline
\end{tabular}
\caption{Experimental values and SM predictions for the observables used in the numerical analysis.}
\label{tbl:obs}
\end{table}
\FloatBarrier
\section{Observables} 
\label{sec:constraints}

In this Section, we analyse the impact of the RPV and DM interactions on low-energy observables. It turns out that the relevant processes are those that receive NP contributions dependent on the unsuppressed couplings $\lambda_{323}$, $\lambda'_{233}$, $\lambda'_{333}$, $\hat{\lambda}_{32}$ and $\hat{\lambda}_{33}$ or at least by the $\epsilon$-suppressed couplings $\lambda'_{223}$ and $\lambda'_{323}$. The respective experimental values and upper bounds are listed in Tbl. \ref{tbl:obs}. 

\subsection{B mesons} 
\label{sec:Bmesons}

\begin{figure}[t]
\centering
\includegraphics[width=7.6cm]{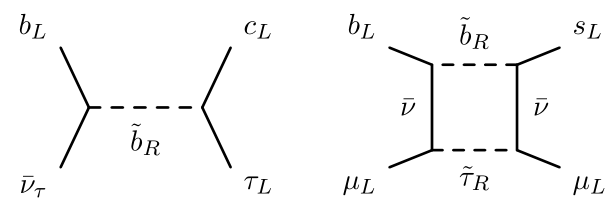}
\caption{The principle tree-level and box diagrams that generate contributions to the charged- and neutral-current B-physics anomalies, respectively.}  
\label{fig:anomalies_diag}
\end{figure}
The processes that involve the B meson are affected solely by RPV interactions. To begin with, one can build four-fermion, semi-leptonic operators that generate contributions to the decay of a bottom quark to second generation quarks from the trilinear terms in \eqref{eq:trilinearlambdaprime} by a tree-level sbottom exchange. Working in the mass eigenbasis for the down-type quarks, the effective Lagrangians read,
\begin{align}
\label{eq:btoclnu}
&\mathcal{L}(b\to c\ell\bar{\nu}_{\ell}) = -\frac{4G_F}{\sqrt{2}}V_{cb}(\delta_{ii'}+\Delta_{ii'}^c)\bar \ell^{i'}_L \gamma^\mu \nu^i_L \bar c_L \gamma_\mu b_L , \\
\label{eq:btosnunu}
&\mathcal{L}(b\to s\nu\bar{\nu}) = \notag \\
&-\frac{4G_F}{\sqrt{2}} \frac{\alpha_{\rm em}}{2 \pi  s_W^2} X_t V_{ts}^* V_{tb} \left( \delta_{ii'}+ X_{ii'}^s \right) \bar \nu^{i'}_L \gamma^\mu \nu^i_L \bar s_L \gamma_\mu b_L ,
\end{align}
where 
\begin{align}
\label{eq:DeltacXs}
&\Delta_{ii'}^c = \sum_{j'=s,b} \frac{\sqrt{2}}{4G_F}\frac{\lambda'_{i33}\lambda_{i'j'3}^{\prime *}}{ 2 m^2_{\tilde b_R}}\frac{V_{cj'}}{V_{cb}}, \\
&X_{ii'}^s = -\frac{\pi s_W^2}{\sqrt{2}G_F \alpha_{\rm em} X_t V_{ts}^* V_{tb}}\left( \frac{\lambda'_{i33}\lambda_{i'23}^{\prime *}}{2 m^2_{\tilde
b_R}} \right),
\end{align}
and $X_t = 1.469 \pm 0.017$ is a SM loop function~\cite{Brod:2010hi}. The processes of interest are the charged-current $b \to c \tau \nu$ anomaly and the $B \to K^{(*)} \nu \bar{\nu}$ decay. The NP effects are probed by the ratios~\cite{Deshpand:2016cpw}:
\begin{align} 
\label{eq:rD}
&r_{D^{(*)}} = \frac{R_{D^{(*)}}}{R_{D^{(*)}}^{\text{SM}}}  = \frac{\left|1+\Delta^c_{33}\right|^2+\left|\Delta^c_{23}\right|^2}{\frac{1}{2}\left(1+\left|1+\Delta^c_{22}\right|^2+\left|\Delta^c_{32}\right|^2\right)} \ \ \ \text{and} \\
\label{eq:RBKnunu}
&R_{B \to K^{(*)} \nu \bar{\nu}} = \frac{ \mathcal B(B \to K^{(*)} \nu \bar{\nu})}{ \mathcal B(B \to K^{(*)} \nu \bar{\nu})_{SM} } \notag \\
&= \sum_{i=\mu,\tau} \frac{1}{3} \left| 1+ X_{ii}^s \right|^2 + \sum_{i\neq i'} \frac{1}{3} \left| X_{ii'}^s \right|^2.
\end{align}
The reported enhancement of $r_{D^{(*)}}$ favours large (resp. small) values for the coupling $\left|\lambda'_{333}\right|$ (resp. $\left|\lambda'_{233}\right|$) and the coupling combination $\lambda'_{333} \lambda_{323}^{\prime *}$ (resp. $\lambda_{233} \lambda_{223}^{\prime *}$), while $R_{B \to K^{(*)} \nu \bar{\nu}}$ sets a strong upper bound on the sum of the same coupling combinations. \footnotetext{The value for $r_{D^{(*)}}$ is calculated according to Eq. \eqref{eq:rD} as the fraction of the weighted average of the $R_D$ and $R_{D^*}$ most recent experimental world average~\cite{Belle:Moriond} over the weighted average of the $R_D^{\text{SM}}$ and $R_{D^*}^{\text{SM}}$ predictions~\cite{Bernlochner:2017jka, Bigi:2017jbd}.} \par
The effective Lagrangian describing the neutral-current $b \to s \ell\bar{\ell}$ decay is parametrized as,
\begin{align}
\label{eq:btosll}
\mathcal{L}(b\to s\ell \bar \ell) &= \frac{4G_F}{\sqrt{2}}\frac{\alpha_{\rm em}}{4 \pi}V_{tb}V_{tb}^*\Big[(C_9^{\ell}+\delta C_9^{\ell})\bar \ell \gamma^\mu \ell \bar s_L \gamma_\mu b_L \notag \\
 &+ (C_{10}^{\ell}+\delta C_{10}^{\ell})\bar \ell \gamma^\mu \gamma_5 \ell \bar s_L \gamma_\mu b_L \notag \\
 &+({C'}_9^{\ell}+\delta {C'}_9^{\ell})\bar \ell \gamma^\mu \ell \bar s_R \gamma_\mu b_R \notag \\
 &\left. +({C'}_{10}^{\ell}+\delta {C'}_{10}^{\ell})\bar \ell \gamma^\mu \gamma_5 \ell \bar s_R \gamma_\mu b_R \right].
\end{align}
In our framework, the operator $\bar \mu^{i'}_L \gamma^\mu \mu^i_L \bar d^k_R \gamma_\mu d^{k'}_R$ is generated at tree-level and the operator $\bar \mu^{i'}_L \gamma^\mu \mu^i_L \bar d^k_L \gamma_\mu d^{k'}_L$ at one-loop level~\cite{Das:2017kfo} giving rise to the correlations $\delta C_{9}^{\mu} = -\delta C_{10}^{\mu}$ and $\delta {C'}_{9}^{\mu} = -\delta {C'}_{10}^{\mu}$. \par 
Let us examine first the tree-level case,
\begin{equation}
\delta {C'}_9^{\mu} = -\delta {C'}_{10}^{\mu} = \frac{\pi \sqrt{2}}{G_F \alpha_e}\frac{1}{V_{tb}V_{ts}^*}\frac{\lambda'_{232} \lambda'_{233}}{4 m_{\tilde{t}_L}^2}.
\end{equation}
The flavour symmetry expectation is $\lambda'_{232} \lambda'_{233} \sim \epsilon_{\bar{F}}^2 \epsilon \approx 0.025$ and hence $\delta C'_9 \approx (8\times 10^6 ~\rm GeV^2) /  m_{\tilde{t}_L}^2$. The operator is disfavoured as an explanation for the anomalies~\cite{Alguero:2019ppt}, which implies that the left-handed stop has to be heavier than a few $\rm TeV$, as already assumed in Sec. \ref{sec:RPV_interactions}. Concretely, the global fit result $\delta {C'}_9^{\mu} = - \delta {C'}_{10}^{\mu} \lesssim 0.12$ is satisfied for $m_{\tilde{t}_L}^2 \gtrsim 8 ~\rm TeV$. We also note, that for this reason the solutions of Refs.~\cite{Das:2017kfo} and~\cite{Earl:2018snx} for the $R_K^{(*)}$ anomaly cannot be applied. \par
Next, the NP effect at one-loop level is
\begin{align}
\label{eq:C9}
\delta C_{9}^{\mu} &= - \delta C_{10}^{\mu} = - \frac{m_t^2}{16 \pi \alpha_{\rm em}}\frac{\left|\lambda'_{233}\right|^2}{m_{\tilde{b}_R}^2} \notag \\
&-\frac{\lambda'_{i33}\lambda_{i23}^{\prime *}\left|\lambda'_{2j3}\right|^2}{64 \sqrt{2} G_F \pi V_{tb}V_{ts}^* \alpha_{\rm em} m_{\tilde{b}_R}^2} \notag \\
& - \frac{\lambda'_{333}\lambda_{323}^{\prime *} \left|\lambda_{323}\right|^2}{64 \sqrt{2} G_F \pi V_{tb}V_{ts}^* \alpha_{\rm em}m_{\tilde{b}_R}^2}\frac{\log \left(m_{\tilde{b}_R}^2/m_{\tilde{\tau}_R}^2\right)}{m_{\tilde{b}_R}^2-m_{\tilde{\tau}_R}^2}.
\end{align}
The first term corresponds to a box diagram with a $W$ boson, a sbottom and two tops in the loop, the second to a box with two sbottoms and two tops and the third to a box with a sbottom, a stau and two tau neutrinos. The first two terms become negligible due to the above-mentioned constraints from the tree-level processes. The third term provides the principal contribution to the anomaly and it is mainly constrained by purely leptonic processes (see next Sec. \ref{sec:taudecays}). \par
Box diagrams with two sbottoms and two tau neutrinos can also affect the $B_s - \bar{B_s}$ mixing~\cite{Earl:2018snx} (see App. \ref{app:DeltaB=2}) and impose similar constraints to the ones from $R_{B \to K^{(*)} \nu \bar{\nu}}$ yet slightly weaker. \par
As a final remark, we point out that the recent analysis in \cite{Aebischer:2019mlg} considers also the likelihood in the space of pairs of WCs. It is suggested that the $\delta C_{9}^{\mu} = -\delta C_{10}^{\mu}$ solution is complemented by a non-vanishing, lepton universal, negative contribution $\delta C_{9}^U$. The statistical significance of this scenario is as of now not overwhelming and we do not take it into account here. If the tendency persists, we have verified that the RPV framework can also accommodate it by the inclusion of loop-diagrams with light left-handed tau sneutrinos around the $\rm TeV$ scale. As a bonus, it is possible to explain partially the anomalous magnetic dipole moment of the muon (see App. \ref{app:g-2}). However, the price to be paid is the abandonment of our flavour symmetry. In particular, the operator $(\bar{b}_R^{\alpha} s_L^{\alpha}) (\bar{b}_L^{\beta} s_R^{\beta})$ is generated by a sneutrino $\tilde{\nu}_L$ tree-level exchange and the corresponding WC, $C_{B_s}^{LR} = \frac{\lambda'_{332}\lambda_{323}^{\prime *}}{2 m_{\tilde{\nu}_L}^2}$ is restricted to be less than $10^{-4} ~ \rm TeV^{-2}$ by the $B_s - \bar{B_s}$ mixing bounds~\cite{Wang:2010vv}. We expect $\lambda'_{323} \sim \epsilon \approx 0.025$ and hence for $m_{\bar{\nu}_L} = 1 ~ \rm TeV$, we get $\lambda'_{332} \lesssim 6 \times 10^{-3}$, which is in contradiction to the flavour symmetry expectation $\lambda'_{332} \sim \epsilon_{\bar{F}} \approx 1$.

\subsection{$Z \to \ell \bar{\ell '}$ coupling and LFV $\tau$ decays} 
\label{sec:taudecays}

\begin{figure}[t]
\centering
\includegraphics[width=7.6cm]{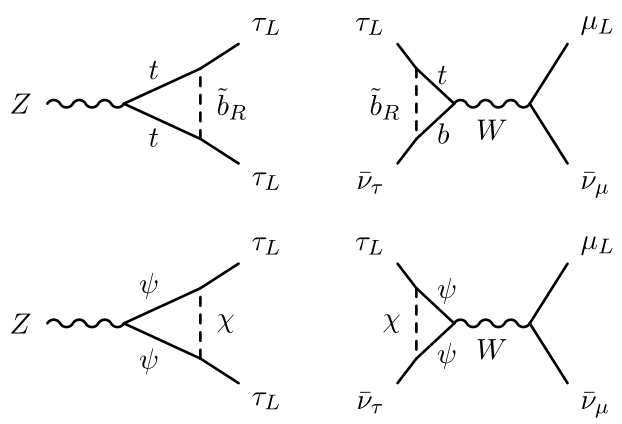}
\caption{The triangle and W-penguin diagrams involving RPV and DM interactions that generate contributions to the $\frac{a_{\tau}}{a_{e}}$ and $R_{\tau}^{\tau/\ell}$ observables, respectively. There are two more diagrams involving RPV interactions which result by exchanging the SM quarks with the sbottom (since it also couples to the gauge bosons) in the loop.}  
\label{fig:Ztau-diag}
\end{figure}
The main obstacles to a combined solution to the anomalies in the generic RPV scenario are the stringent constraints from the $Z$ boson decay to a dilepton pair and the charged-current leptonic tau decay to a lepton and neutrinos~\cite{Feruglio:2016gvd}~\cite{Feruglio:2017rjo}. On the one hand, the triangle diagrams involving the sbottom and the top generate a positive contribution to the leptonic $Z$ coupling that saturates the experimental bound already for order $\mathcal{O}(1)$ values of $\lambda'_{333}$. On the other hand, the tree-level stau exchange implies an unacceptably large shift of the $\tau \to l \nu \bar{\nu}$ decay rate~\cite{Kao:2009fg}, unless the stau mass is of order $10 \ \rm TeV$. The third term in \eqref{eq:C9} becomes also negligible and thus the $R_{K^{(*)}}$ anomaly remains unresolved. Moreover, no cancellation between the positive tree-level amplitude and the negative $W$ penguin diagrams involving the sbottom and the top can be invoked due to the previous bounds on $\lambda'_{333}$ from the $Z$ coupling. \par
We stress that the necessity of a cancellation between the tree-level purely leptonic interaction and the leptoquark RGE effects appearing at one-loop level is already anticipated by the EFT analysis based on $U(2)$ flavour symmetries~\cite{Bordone:2017anc}. Motivated by this analysis, we examine how the DM interactions may alter the above conclusions. Let us write down the relevant observables, i.e. the ratio of $Z$ boson axial-vector coupling,
\begin{align}
\frac{a_{\tau}}{a_{e}} = 1 - \frac{3 m_t^2}{16 \pi^2} \frac{\left|\lambda'_{333}\right|^2}{m_{\tilde{b}_R}^2} \left(\log \left(\frac{m_{\tilde{b}_R}^2}{m_t^2}\right) - 1 \right) \notag \\
+(1-4s_W^2)\frac{|\hat{\lambda}_{33}|^2}{16 \pi^2} \left(\frac{m_{\chi}^2}{m_{\psi}^2} \log \left(\frac{m_{\chi}^2}{m_{\psi}^2}\right) + 1 \right)
\end{align}
and the following ratio of leptonic decay branching ratios at leading order,
\begin{align}
\label{eq:Rtauldef}
R_{\tau}^{\tau/\ell} &= \frac{ \mathcal B(\tau \to \ell  \nu \bar{\nu})_{\text{exp}} / \mathcal B(\tau \to \ell  \nu \bar{\nu})_{\text{SM}}}{ \mathcal B(\mu \to e \nu \bar{\nu})_{\text{exp}} / \mathcal B(\mu \to e \nu \bar{\nu})_{\text{SM}}}  \simeq  1 + \frac{\sqrt{2}}{4G_F}\frac{\left|\lambda_{323}\right|^2}{m_{\tilde{\tau}_R}^2} \notag \\
&-\frac{3 m_t^2}{16 \pi^2}\frac{\left|\lambda'_{333}\right|^2}{m_{\tilde{b}_R}^2} \left(\log{\left(\frac{m_{\tilde{b}_R}^2}{m_t^2}\right)}-\frac{1}{2}\right) \notag \\
&-\frac{\hat{\lambda}_{32}\hat{\lambda}_{33}^*}{8 \pi^2}\left(\frac{m_{\chi}^2}{m_{\psi}^2} \log \left(\frac{m_{\chi}^2}{m_{\psi}^2}\right) + 1 \right).
\end{align}
We observe that the inclusion of the new scalar particle $\chi$ together with the mediator $\psi$ can potentially improve the situation. Not only, do they enter a triangle diagram that contributes to $\frac{a_{\tau}}{a_{e}}$ with an opposite sign to the RPV term, but also a $W$ penguin diagram that can facilitate further the cancellation in $R_{\tau}^{\tau/\ell}$. It should be noted that, if the DM particle were the Dirac fermion in $X_3$ the contribution to the $Z$ coupling would be negligible. \par
In the presence of the DM interactions, the effects on two more LFV $\tau$ decays should be taken into account. In particular, $Z$ penguin diagrams with either the sbottom and the top~\cite{Earl:2018snx} or the $\chi$ and the $\psi$ in the loop may pose a problem with regard to the SM forbidden $\tau \to \mu \mu \bar{\mu}$ decay\footnote{There are also box diagrams~\cite{Earl:2018snx} and photon penguin~\cite{deGouvea:2000cf} diagrams that contribute to this decay but we have explicitly checked that they are subleading.} and triangle diagrams with the same particles~\cite{deGouvea:2000cf}~\cite{Klasen:2016vgl} regarding the radiative $\tau \to \mu \gamma$ decay (see App. \ref{app:LFVtaudec}). 

\subsection{Relic abundance and Dark Matter direct detection} 
\label{sec:DMpheno}

The DM particle couples directly to leptons and therefore the main self-annihilation channels are $\bar{\chi} \chi \to \bar{\ell} \ell$ and $\bar{\chi} \chi \to \bar{\nu}_{\ell} \nu_{\ell}$ with $\ell = \mu, \tau$. Because $\chi$ is a scalar particle, the effective cross-section is p-wave suppressed~\cite{Bai:2014osa},
\begin{equation}
\label{eq:chiannihilation}
\frac{1}{2} \left\langle \sigma v \right\rangle = \frac{1}{2} \left[ \frac{(|\hat{\lambda}_{32}|^2+|\hat{\lambda}_{33}|^2) m_{\chi}^2}{48 \pi (m_{\psi}^2+m_{\chi}^2)^2} v^2 \right] \equiv p v^2,
\end{equation}
where $v \sim 3$ is the DM velocity at freeze-out. The relic abundance can be estimated following the calculation in Ref.~\cite{Bai:2013iqa}. Co-annihilation effects are irrelevant since the rest of the DM generations are taken to be heavier. \par 
Concerning direct detection, the dominant contribution for DM scattering off nucleons is generated via a penguin diagram involving leptons and the charged mediator in the loop and a virtual photon exchange. The respective dimension-6 operator is called charge-radius operator,
\begin{equation}
\label{eq:Lcharge-radius}
\mathcal{L}_{\text{charge-radius}} = i b_{\chi} \partial_{\mu} \chi^* \partial_{\nu} \chi F^{\mu \nu},
\end{equation}
where
\begin{equation}
\label{eq:bcharge-radius}
b_{\chi} = \sum_{\ell=\mu,\tau} \frac{|\hat{\lambda}_{3\ell}|^2 e}{16 \pi^2 m_{\psi}^2} \left( 1 - \frac{2}{3} \log \left(\frac{m_{\ell}^2}{m_{\psi}^2}\right) \right).
\end{equation}
The spin-independent DM-nucleus differential scattering cross section is then~\cite{Bai:2014osa},
\begin{equation}
\label{eq:DM-Ndiffsigma}
\frac{d \sigma}{d E_R} = \frac{Z^2 e^2 b_{\chi}^2 m_T}{16 \pi v^2} F_E^2(q^2).
\end{equation}
Here, $E_R = \left|\vec{q}\right|^2 / 2 m_T$ is the recoil energy; $v$ is the DM velocity in the lab frame; $m_T$ is the mass, $Z$ the atomic number and $F_E(q^2)$ the electric form factor of the target nucleus~\cite{Banks:2010eh}. Eq. \eqref{eq:DM-Ndiffsigma} has the same $E_R$- and $v^2$-dependence as the ordinary spin-independent cross section for a contact interaction, we can directly map the latest, most stringent exclusion limits, as published by the XENON1T collaboration~\cite{Aprile:2018dbl}, onto limits on the parameter space of the current model. Since, the virtual photon couples only to protons inside the nucleus, a rescaling with $Z^2 / A^2$ is necessary in order to account for the resulting isospin violation. We also derive projections for XENONnT, which is the next upgrade step of XENON1T and will increase the target mass to $5.9 t$, by assuming the same efficiency profile. \par
Lastly, we mention that indirect detection signals of scalar DM are too small to be observed due to the p-wave suppression of the cross-section~\cite{Kavanagh:2018xeh}.

\subsection{Collider searches} 
\label{sec:collider}

We highlight the basic aspects of the high-$p_T$ searches for the new particles. The direct comparison with analyses that study the analogous EFT operators offers the possibility to set present and future bounds for their masses and interactions.
\begin{itemize}
	\item $\boldsymbol{\tilde{b}_R.}$~~Crossing symmetry dictates that an explanation for $R_{D^{(*)}}$ is unambiguously connected to the scattering $b c \to \tau \bar{\nu}$. The data from the mono-tau signature $pp \to \tau_h X + \text{MET}$ at LHC set upper bounds on the WC of the effective operator generated by a scalar leptoquark $S_1 \sim (\bar{\mathbf{3}}, \mathbf{1})_{1/3}$ exchange~\cite{Altmannshofer:2017poe}~\cite{Greljo:2018tzh}, which for our case translate into $\Delta_{33}^c < 0.32$ (see Eq. \eqref{eq:DeltacXs}). The prognosis for the HL-LHC ($3000~\rm fb^{-1}$) gives $\Delta_{33}^c < 0.1$.
		\item $\boldsymbol{\tilde{\tau}_R.}$~~Depending on the nature of the LSP, the possible signatures involving the stau are classified in Ref.~\cite{Dercks:2017lfq}. The best limits in RPV searches using simplified models yield $m_{\tilde{\tau}_R} > 300 \rm ~ GeV$.   
		\item $\boldsymbol{\chi \ \& \ \psi.}$~~The mediator $\psi$ can be pair-produced at LHC via Drell-Yan processes, i.e. a $Z$ boson or photon exchange, with di-lepton plus missing energy signature $l_i^+ l_j^- + \text{MET}$. The cross sections is larger than that of fermion DM models, because there are more helicity states for the fermionic mediator. If the leptons have the same flavour, one can recast slepton searches and extrapolate the limits: $m_{\chi} > 300 ~ \rm GeV$ and $m_{\psi} > 500 ~ \rm GeV$~\cite{Chang:2014tea}. Likewise, $\psi \bar{\psi}$ direct production is possible at LEP, but the bounds are even weaker~\cite{Kavanagh:2018xeh}. Other channels accessible at future muon colliders that are relevant to our setup could be $\mu^+ \mu^- \to \chi \bar{\chi} \gamma$ at tree-level with mono-photon search signature and $\mu^+ \mu^- \to l_i^+ l_j^- \gamma$ at one-loop level with multi-flavour lepton final state (in analogy to the discussion in Ref.~\cite{Chen:2015jkt} for LEP).
\end{itemize}

\section{Phenomenological Analysis} 
\label{sec:fit}

The preferred region of the parameter space is determined by performing a minimization of the $\chi^2$-distribution composed by the observables of the previous Section \ref{sec:constraints}. We use the experimental data of Tbl. \ref{tbl:obs}, the XENON1T exclusion limits~\cite{Aprile:2018dbl}, the mass lower bounds set by collider searches (see Sec. \ref{sec:collider}) and the necessary SM input~\cite{Patrignani:2016xqp}. The values of the couplings are taken to be less than $\sqrt{4 \pi}$, while the complex phases are not directly constrained. As a matter of fact, what enters the expressions of the various observables are the absolute values of single couplings or coupling combinations. The only parameter that may potentially probe directly the size of CP-violating phases is $\phi_{B_s}$ yet the overall flavour suppression excludes this possibility at the moment\footnote{We estimate the impact of complex phases according to Eq. \eqref{eq:phiBs} and get $\phi_{B_s} \in [-0.4^{\circ},0.4^{\circ}]$, which is well within the current experimental limits.}. \par
\begin{figure}[t!]
\centering
\includegraphics[width=8.6cm]{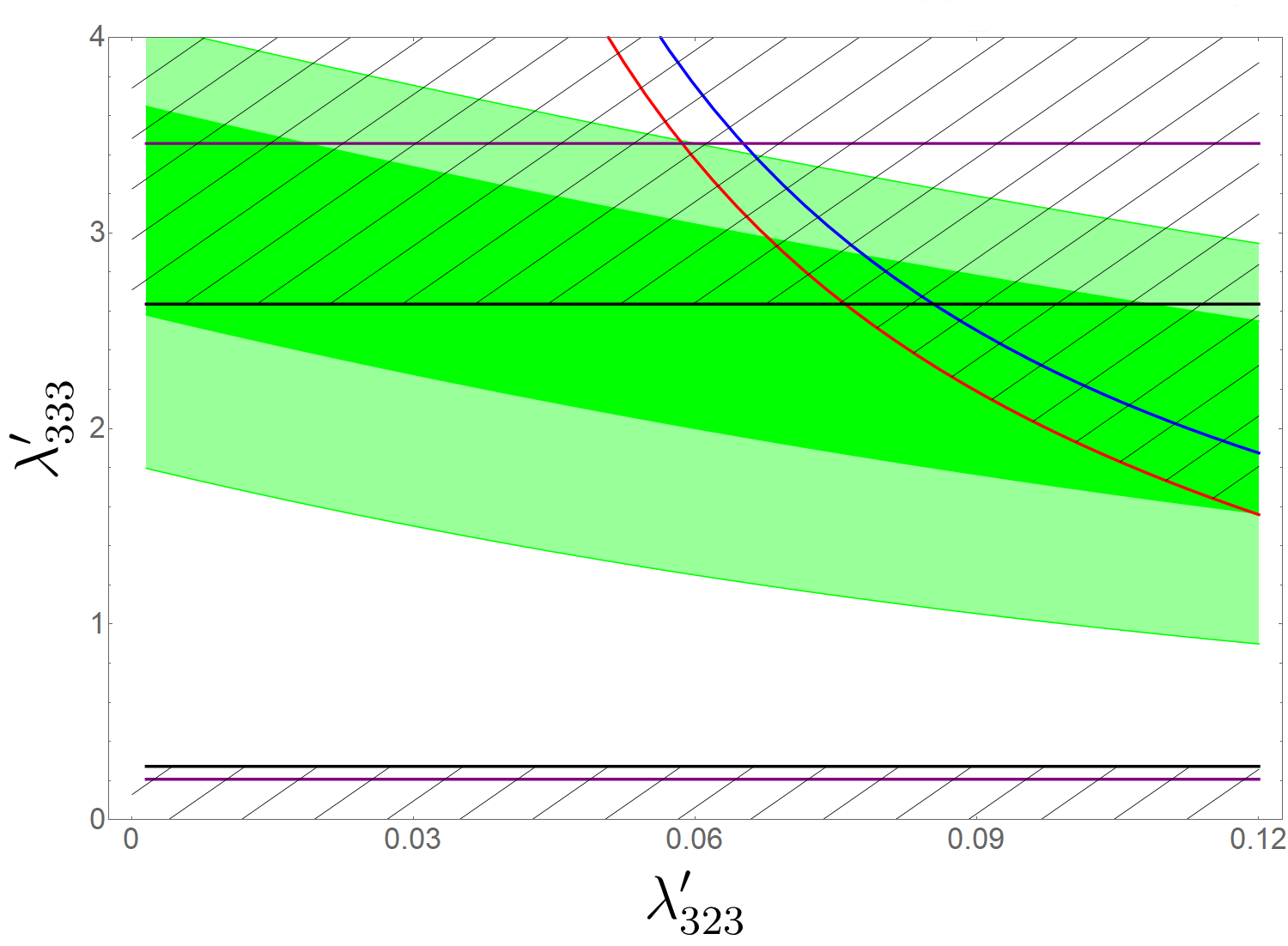} \\[5pt]
\includegraphics[width=8.6cm]{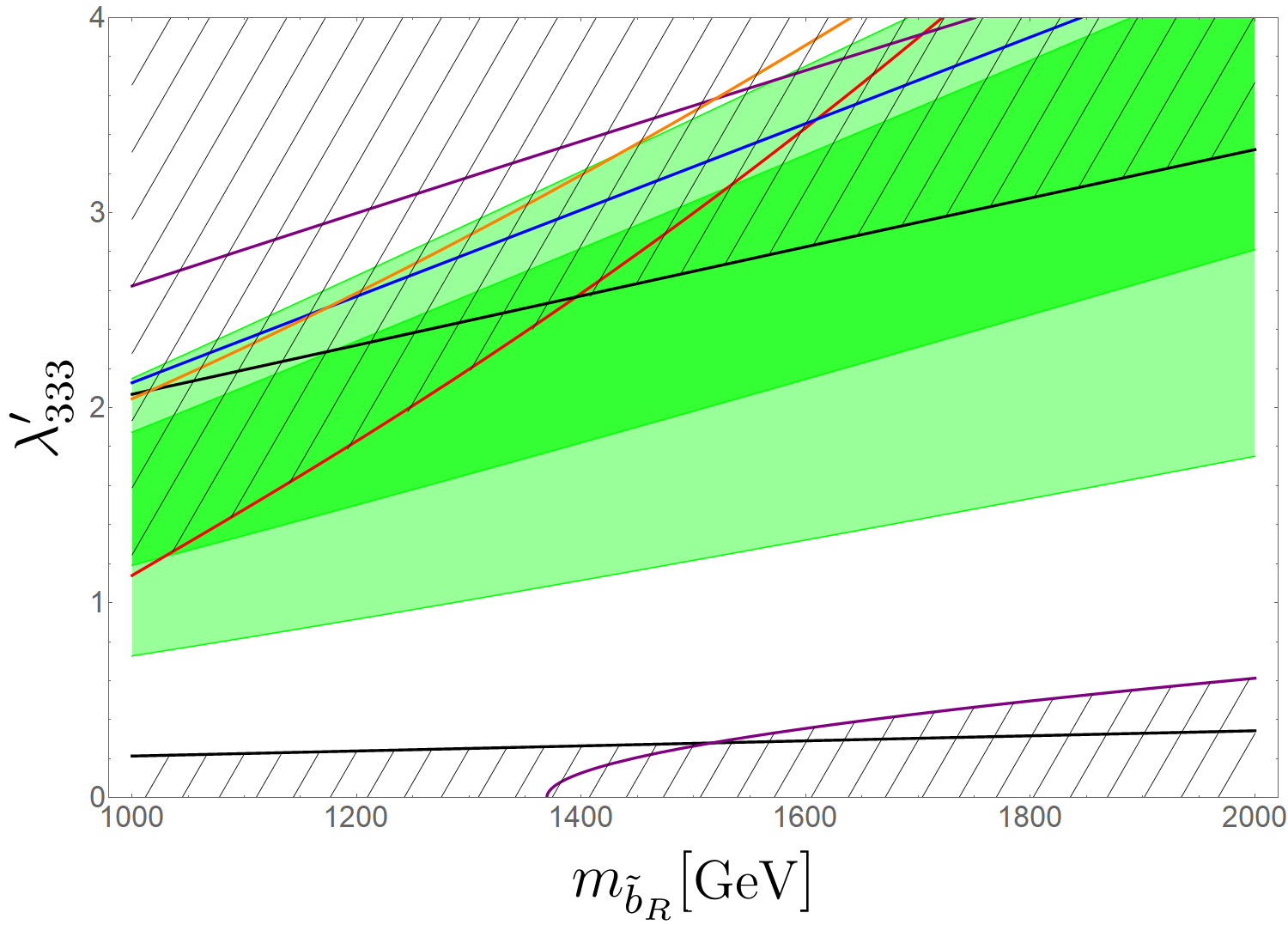} \\[5pt]
\includegraphics[width=8.6cm]{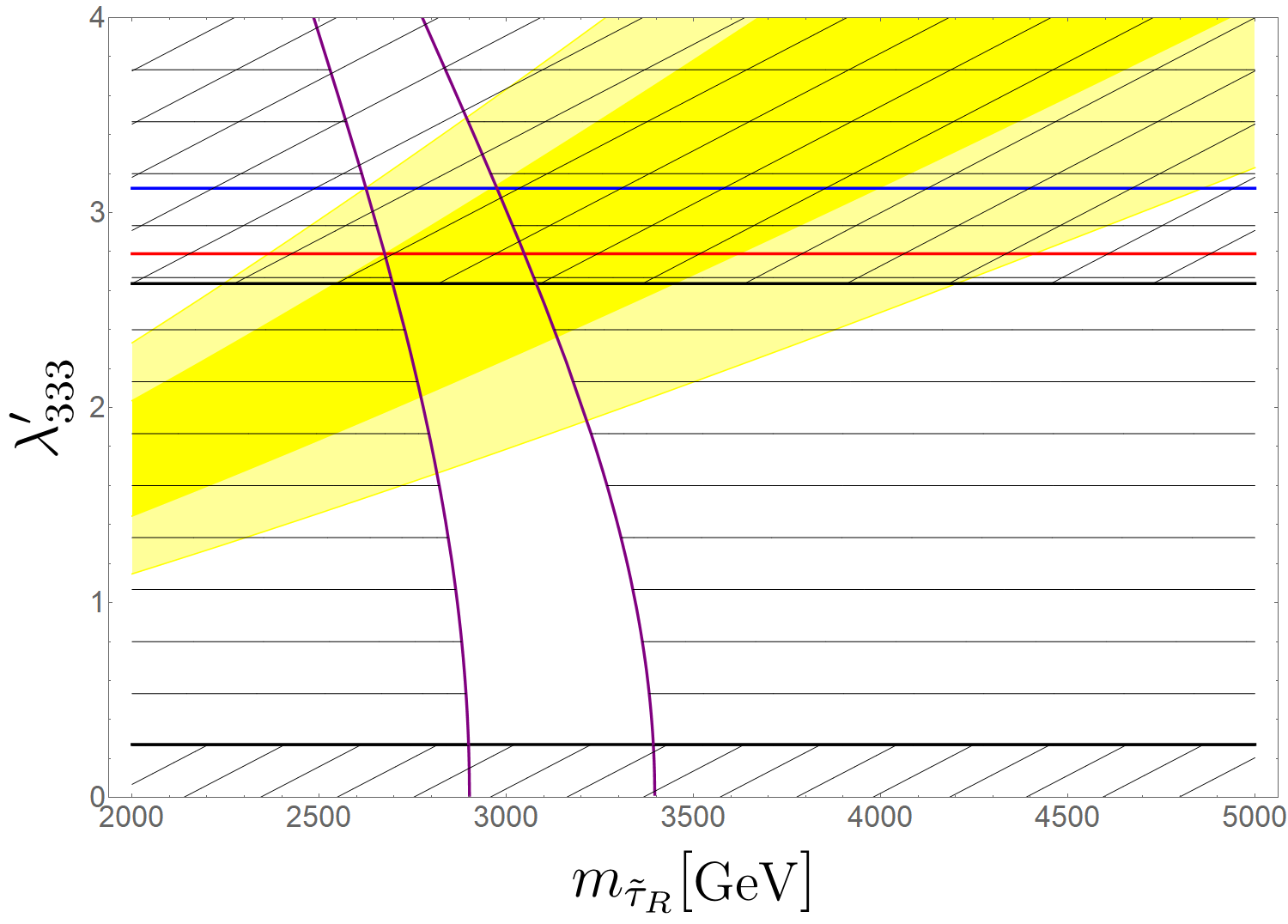}
\caption{RPV parameter space compatible with a solution for $r_{D^{(*)}}$ (green) and $\delta C_9^{\mu}$ (yellow) at $1 \sigma$ and $2 \sigma$ around the best-fit point. The hatched region is excluded at $2 \sigma$ due to the following constraints: $R_{B \to K^{(*)} \nu \bar{\nu}}$ (red), $C_{B_s}$ (blue), $\frac{a_{\tau}}{a_{e}}$ (black), $R_{\tau}^{\tau/\ell }$ (purple), $\mathcal B(\tau \to \mu \mu \bar{\mu})$ (orange).}
\label{fig:RPV2Dplanes}
\end{figure}

The best-fit point is presented in Tbl. \ref{tbl:bfp}. The improvement of the total $\chi^2$ with respect to the SM limit $\chi^2 (x_{\text{SM}} ) - \chi^2 (x_{\text{BF}} ) \simeq 33 - 5 = 27$ reflects the resolution of the anomalies. This is also evident in the Figures \ref{fig:RPV2Dplanes}, where the $68\%$ CL and $95\%$ CL regions of the $r_{D^{(*)}}$ and $\delta C_9^{\mu}$ observables in the $(\lambda'_{333},\lambda'_{323})$, $(\lambda'_{333},m_{\tilde{b}_R})$ and $(\lambda'_{333},m_{\tilde{\tau}_R})$ planes are shown together with the $2\sigma$ exclusion contours from the other low-energy observables.
\begin{table}
\includegraphics[width=\linewidth]{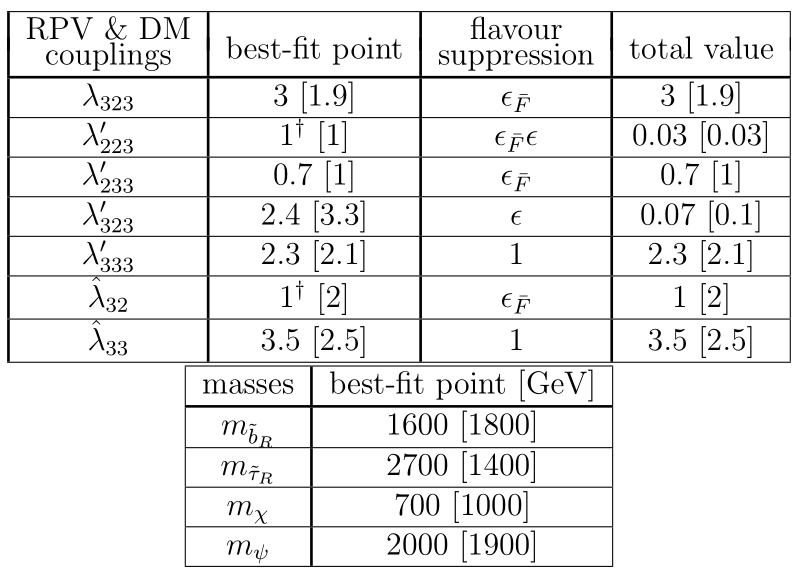}
\caption{The best-fit point for the couplings along with the flavour suppression factors and the particle masses. The entries marked with a dagger are not determined by the fit, but are rather benchmark points (see text). The values in the brackets denote a fit-point that allows for a smaller stau mass still within the $1 \sigma$ region for the anomalies .}
\label{tbl:bfp}
\end{table}
\begin{figure}[t]
\centering
\includegraphics[width=8.6cm]{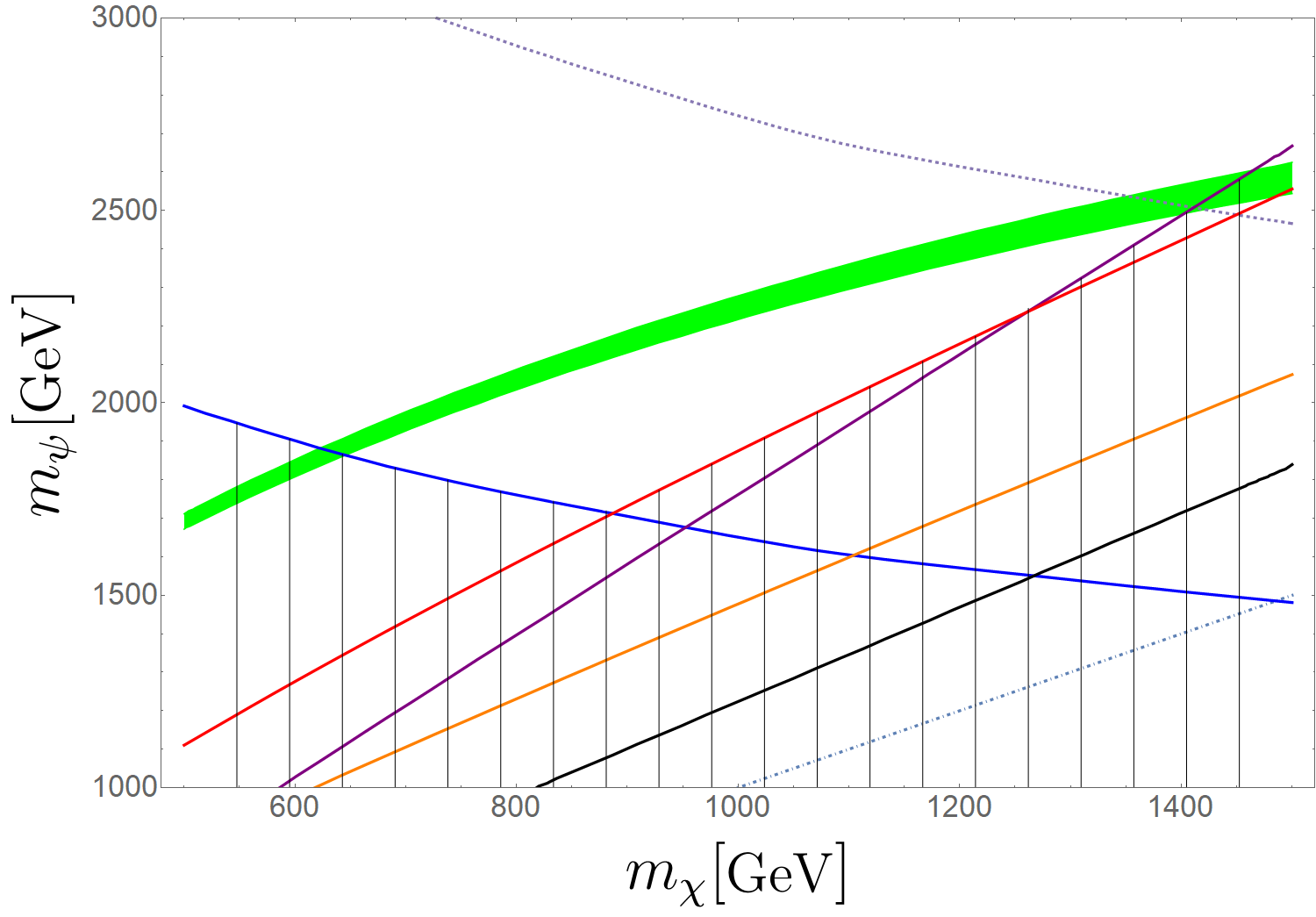}
\caption{DM parameter space compatible with the correct relic abundance (green) at $1 \sigma$ around the best-fit point. The hatched region is excluded at $2 \sigma$ due to the following constraints: XENON1T exclusion limits (blue), $\frac{a_{\tau}}{a_{e}}$ (black), $R_{\tau}^{\tau/\ell }$ (purple), $\mathcal B(\tau \to \mu \gamma)$ (red), $\mathcal B(\tau \to \mu \mu \bar{\mu})$ (orange). The dashed line represents the prospective XENONnT exclusion limits.}  
\label{fig:DM2Dplane}
\end{figure}
We observe that a large fraction of the parameter space becomes available due the cancellation mechanism in Sec. \ref{sec:taudecays}. In particular, the coupling $\lambda'_{333}$ reaches now higher values compared with the generic RPV scenario~\cite{Altmannshofer:2017poe} and the bounds on $B_s - \bar{B_s}$ mixing and $R_{B \to K^{(*)} \nu \bar{\nu}}$ are still satisfied thanks to the flavour suppression in $\lambda'_{323}$. Moreover, the stau can be relatively light enhancing adequately the last term in Eq. \eqref{eq:C9}. Both $R_{D^{(*)}}$ and $R_{K^{(*)}}$ are then in good agreement with the present central values and the mass spectrum is more natural with all the right-handed superparters, $\chi$ and $\psi$ within the same mass range of a few $\rm TeV$. However, due to the sensitivity of $R_{\tau}^{\tau/\ell}$ to the cancellation, $R_{K^{(*)}}$ can only be accommodated in a concrete band of the $(\lambda'_{333},m_{\tilde{\tau}_R})$ plane, which for this solution is located around a stau mass of approximately $2 \ \rm TeV$. By varying the coupling $\lambda_{323}$ and the DM parameters, the position of the band can be moved to the left as low as $\sim 1.4 ~ \rm TeV$ before abandoning the $1 \sigma$ region for $\delta C_9^{\mu}$ (see the values in brackets in Tbl. \ref{tbl:bfp}). As for the couplings $\lambda'_{223}$ and $\lambda'_{233}$, the first is flavour suppressed and does not affect the fit and the latter is primarily constrained by the bounds on LFV $\tau$ decays (see Eqs. \eqref{eq:Ltauto3mu}-\eqref{eq:Btautomugamma}). \par
Regarding the DM interactions, we notice that the coupling $\hat{\lambda}_{32}$ appears only in a product combination with the coupling $\hat{\lambda}_{33}$ in the observables of Sec. \ref{sec:taudecays} and the single-coupling bounds from e.g. $\frac{a_{\mu}}{a_{e}}$, are satisfied for any order $\mathcal{O}(1)$ value. Consequently, it is convenient to use the benchmark $\hat{\lambda}_{32} = 1$ and include only $\hat{\lambda}_{33}$ in the fit. The region in the $(m_{\psi},m_{\chi})$ plane that can give rise to the correct relic abundance is shown in Figure \ref{fig:DM2Dplane}. The present direct detection and LFV bounds allow for a DM mass $700 ~ \rm GeV < m_{\chi} < 1.5 ~ \rm TeV$ and we have explicitly verified that the cancellation mechanism is applicable for the whole mass range by readjusting the rest of the parameters. One may additionally quantify the percentage of cancellation needed between the NP terms in $\frac{a_{\tau}}{a_{e}}$ and $R_{\tau}^{\tau/\ell}$ and find that it amounts to approximately $40 \%$ and $10 \%$, respectively. This is a rather mild condition and it can be fulfilled independently of the relative sign of the couplings. Note that the cancellation in $R_{\tau}^{\tau/\ell}$ is exactly of the same order as the one predicted by the general EFT analysis~\cite{Bordone:2017anc}. \par
As a final note, we comment on the testability of the model by future experiments. With the most recent world average for the charged-current anomalies~\cite{Belle:Moriond}, we calculate $\Delta_{33}^c \approx 0.065$. The `no-loose theorem' of Ref.~\cite{Greljo:2018tzh} no longer applies and there is still an open window for the leptoquark explanation of $R_{D^{(*)}}$, even if there is no discovery after the HL-LHC phase. The situation for the stau and $\chi$ is even more inconclusive, since the masses for these particles appear to be out of reach of the LHC. On the other hand, DM direct detection looks much more promising. We see that the bulk of the parameter space for the chosen benchmark will be probed by XENONnT (and other experiments that aim at similar exposure). If the anomalies are univocally confirmed as NP signals, it is worth analysing the sensitivity of the next generation of colliders to the heavy particles predicted by the model. 

\section{Conclusions}

In this paper, we further investigate the possibility to provide a simultaneous explanation for the B-physics anomalies within the framework of RPV interactions controlled by an approximate $U(2)^2$ flavour symmetry. As pathfinder we use the observation, established by the general EFT analysis, that a destructive amplitude interference in purely leptonic interactions may cure the tensions with specific LFV observables that put limitations on the viability of the scalar leptoquark solution. Furthermore, we assume that the violation of R-parity necessitates the consideration of a new hidden sector that contains a DM candidate particle and that this sector is also charged under the same flavour symmetry. We show that in this case, from the plethora of flavoured DM models, one particular model that features a leptophilic, scalar DM particle is uniquely singled out and favoured from the low-energy fit. It is ensured that all newly introduced particles, interactions and symmetry breaking patterns are in accordance with the spirit of gauge coupling unification. If the sbottom is the LSP, SUSY may remain elusive for the LHC, whereas experimental validation of the proposed DM interaction by direct detection searches can be expected in the near future.

\section{Acknowledgements}
We thank Gino Isidori, Javier Fuentes-Martin and Michael Baker for useful  discussions and comments on the manuscript. This research was supported in part by the Swiss National Science Foundation (SNF) under contract 200021-159720.

\appendix
\section{Additional flavour constraints}

\subsection{$\Delta B = 2$}
\label{app:DeltaB=2}

It is useful to define the effective Hamiltonian, 
\begin{equation}
\label{eq:DeltaB2}
\mathcal{H}(\Delta B = 2)^{\text{SM(NP)}} = - C_{B_s}^{\text{SM(NP)}} (\bar{b}_L \gamma^{\mu} s_L) (\bar{b}_L \gamma_{\mu} s_L),
\end{equation}
where
\begin{align}
\label{eq:CVLL}
&C_{B_s}^{\text{SM}} = -\frac{g^4}{128 \pi^2 m_W^2} (V_{tb} V_{ts}^*)^2 S_0(m_t^2/m_W^2), \notag \\
&C_{B_s}^{\text{NP}} = \frac{\lambda'_{i33}\lambda_{i23}^{\prime *}\lambda'_{j33}\lambda_{j23}^{\prime *}}{128 \pi^2 m^2_{\tilde b_R}}.
\end{align}
and $S_0(x) = \frac{x (4-11x+x^2)}{4(1-x)^2}-\frac{3 x^3 \log(x)}{2(1-x)^3}$. \par
We compare the ratio,
\begin{equation}
\label{eq:CBs}
C_{B_s} = \frac{\left|\bra{B_s^0} \mathcal{H}(\Delta B = 2)^{SM+NP} \ket{\bar{B}_s^0}\right|}{\left|\bra{B_s^0} \mathcal{H}(\Delta B = 2)^{SM} \ket{\bar{B}_s^0}\right|} = \left| 1 + \frac{C_{B_s}^{\text{NP}}}{C_{B_s}^{\text{SM}}} \right|,
\end{equation}
and the CP-violating phase,
\begin{equation}
\label{eq:phiBs}
\phi_{B_s} = \frac{1}{2} \text{Arg} \left( 1 + \frac{C_{B_s}^{\text{NP}}}{C_{B_s}^{\text{SM}}} \right) 
\end{equation}
with the measured values.

\subsection{$(g-2)_{\mu}$}
\label{app:g-2}

The effective Lagrangian that describes the muon anomalous magnetic dipole moment is,
\begin{equation}
\label{eq:Lg2}
\mathcal{L}(\mu \to \mu \gamma) = \frac{e}{4 m_{\mu}} a_{\mu} \bar{\mu} \sigma_{\alpha \beta} \mu F^{\alpha \beta}
\end{equation}
and the generic expression for leptonic RPV interactions~\cite{Chakraborty:2015bsk} is,
\begin{equation}
\label{eq:alphamu}
a_{\mu} = \frac{m_{\mu}^2}{96 \pi^2} \left|\lambda_{323}\right|^2 \left( \frac{2}{m_{\tilde{\nu}_L}^2} - \frac{1}{m_{\tilde{\tau}_R}^2} \right).
\end{equation}
The discrepancy between the SM prediction and the experiment $\Delta a_{\mu} = a_{\mu}^{\text{exp}} - a_{\mu}^{\text{exp}} = (29.3 ~ \pm ~ 9.0)\times 10^{-10}$ requires a positive contribution that is provided by the first term in the above equation.

\subsection{$\tau \to \mu \mu \bar{\mu}$ and $\tau \to \mu \gamma$}
\label{app:LFVtaudec}

The effective Lagrangian and effective amplitude for these processes can be expressed as,
\begin{align}
\label{eq:Ltauto3mu}
\mathcal{L}(\tau \to \mu \mu \bar{\mu}) = &- g_{LL}^{3\mu} (\bar{\tau}_L \gamma^{\mu} \mu_L)(\bar{\mu}_L \gamma_{\mu} \mu_L) \notag \\
&-g_{LR}^{3\mu} (\bar{\tau}_L \gamma^{\mu} \mu_L)(\bar{\mu}_R \gamma_{\mu} \mu_R), \\
\label{eq:Mtautomugamma}
\mathcal{M}(\tau \to \mu \gamma) &= - e\frac{m_{\tau}}{2}\epsilon^{\alpha} \bar{u}_\mu [i \sigma_{\beta \alpha} q^{\beta}(a_R P_L + a_L P_R)] u_{\tau},
\end{align}
We define,
\begin{align}
\label{eq:gtauto3mu}
g_{LL(LR)}^{3\mu} &= \frac{4 \pi \alpha_{\rm em} \kappa_{L(R)}}{c_W^2 s_W^2 m_Z^2} \left[ \frac{3 m_t^2}{32 \pi^2} \frac{\lambda'_{233} \lambda_{333}^{\prime *}}{m_{\tilde{b}_R}^2} \left( \log \left( \frac{m_{\tilde{b}_R}^2}{m_t^2}\right) -1 \right) \right. \notag \\
&- \left. (1-4s_W^2)\frac{|\hat{\lambda}_{33}|^2}{32 \pi^2} \left(\frac{m_{\chi}^2}{m_{\psi}^2} \log \left(\frac{m_{\chi}^2}{m_{\psi}^2}\right) + 1 \right) \right].
\end{align}
where $\kappa_L = -\frac{1}{2} + s_W^2$, $\kappa_R = s_W^2$ and
\begin{equation}
\label{eq:atautomugamma}
a_L = -\frac{\hat{\lambda}_{32}\hat{\lambda}_{33}^*}{192 \pi^2 m_{\psi}^2} G(m_{\chi}^2/m_{\psi}^2) - \frac{\lambda'_{2i3}\lambda_{3i3}^{\prime *}}{64 \pi^2 m^2_{\tilde b_R}},  \ \ \ \ a_R = 0,
\end{equation}
where $G(x) = \frac{1}{(1-x)^4}(2-3x-6x^2+x^3+6x \log (x))$.
The branching ratios~\cite{Dassinger:2007ru},
\begin{align}
\label{eq:Btauto3mu}
&\mathcal B(\tau \to \mu \mu \bar{\mu}) = \frac{1}{G_F^2} \Bigg[ \frac{\left|g_{LL}^{3\mu}\right|^2 + \left|g_{LR}^{3\mu}\right|^2}{4} \notag \\
&+\left( \log \frac{m_{\tau}^2}{m_{\mu}^2} - \frac{11}{4} \right) (4 \pi \alpha_{\rm em})^2 \left|a_{L}\right|^2 \notag \\
&+\Re \left(a_{L}^* g_{LL}^{3\mu} + \frac{1}{2} a_{L}^* g_{LR}^{3\mu}\right) (4 \pi \alpha_{\rm em}) \Bigg] \mathcal B(\tau \to \mu \bar{\nu}_{\mu} \nu_{\tau}) \\
\label{eq:Btautomugamma}
&\mathcal B(\tau \to \mu \gamma) = \frac{48 \pi^3 \alpha_{\rm em}}{G_F^2} (\left|a_{L}\right|^2 + \left|a_{R}\right|^2) \mathcal B(\tau \to \mu \bar{\nu}_{\mu} \nu_{\tau}),
\end{align}
must then comply with the respective experimental upper bounds.

\end{document}